\documentclass{article}
\usepackage{graphicx}
\usepackage{graphics}
\usepackage{amsmath}
\usepackage{geometry}
\usepackage{authblk}

\title{On-chip interrogator based on Fourier transform spectroscopy}

\author[1]{ Fellipe Grillo Peternella\thanks{Corresponding Author: F.GrilloPeternella@tudelft.nl}}
\author[2]{ Thomas Esselink}
\author[1]{ Bas Dorsman}
\author[3]{Peter Harmsma}
\author[1]{Roland C. Horsten}
\author[1]{Thim Zuidwijk}
\author[1]{H. Paul Urbach}
\author[1]{Aur\`{e}le J. L. Adam}
\affil[1]{Optics Group, ImPhys department, TU Delft, Lorentzweg 1, 2628 CJ Delft, The Netherlands}
\affil[2]{The Hague University of Applied Sciences, Rotterdamseweg 137, 2628 AL Delft, The Netherlands}
\affil[3]{TNO, Stieltjesweg 1, 2628 CK Delft, The Netherlands}

\date{}

\begin{document}

\maketitle

\begin{abstract}
In this paper, the design and the characterization of a novel interrogator based on integrated Fourier transform (FT) spectroscopy is presented. To the best of our knowledge, this is the first integrated FT spectrometer used for the interrogation of photonic sensors. It consists of a planar spatial heterodyne spectrometer, which is implemented using an array of Mach-Zehnder interferometers (MZIs) with different optical path differences. Each MZI employs a 3$\times$3 multi-mode interferometer, allowing the retrieval of the complex Fourier coefficients. We derive a system of non-linear equations whose solution, which is obtained numerically from Newton's method, gives the modulation of the sensor's resonances as a function of time. By taking one of the sensors as a reference, to which no external excitation is applied and its temperature is kept constant, about 92$\%$ of the thermal induced phase drift of the integrated MZIs has  been compensated. The minimum modulation amplitude that is obtained experimentally is 400 fm, which is more than two orders of magnitude smaller than the FT spectrometer resolution.
\end{abstract} 

\section{Introduction}
Photonic based sensors find nowadays a wide range of applications. Acoustic and ultrasound sensors \cite{Peternella2017,Zhang2015}, pressure sensors \cite{Hallynck2012}, biochemical and gas sensors \cite{Vos2009,Hou2017} are examples of sensors based on optical technology. They are low cost, immune to electromagnetic radiation, and operate under a wide range of temperatures. In this paper, we focus our attention on photonic sensors whose transmission or reflection spectra have a peak (or dip) in their lineshape.  Examples are sensors based on fiber Bragg gratings (FBGs)\cite{Hou2017,Liang2018} or on integrated ring resonators \cite{Peternella2017,Zhang2015,Vos2009}. For these sensors, it is possible to build large and multi-purpose sensor arrays by wavelength multiplexing the spectrum of the sensors \cite{Liang2018,Kersey1996}.

The photonic sensors mentioned above are designed in such a way that the signal to be sensed modulates the sensor's resonance wavelength. Interrogation is the technique of demodulating and demultiplexing the response of an array of photonic sensors.  Different methods have been proposed in the past. A common approach is to measure the spectrum of the sensor array using a dispersive spectrometer such as an arrayed waveguide grating (AWG)\cite{Ongqiang2017,Pustakhod2016,Yebo2011} or an echelle grating\cite{Guo2013}. Their sensitivity to the external excitation depends on the spectral resolution of the spectrometer; higher resolution comes at the price of a larger footprint.
Another approach is edge filtering, where the output spectra of the photonic sensors is conveyed to an optical filter whose transfer function is linear within certain range. As the spectrum of the sensor shifts due to the sensing signal, the filter converts the resonance wavelength modulation into power modulation which can be obtained by a photodetector. The main drawback is that a high sensitivity may compromise the wavelength operation range\cite{Tiwari2013}. \textit{Passaro et al} \cite{Passaro2012} reports the spectral scanning as a possible solution, which features a high sensitivity and a large wavelength operation range. On the other hand, most of these interrogators are based on thermal tuning which limits their interrogation speed to a few kHz. Another approach for interrogation is to use passive interferometers such as  Mach-Zehnder interferometers. In combination with a demultiplexing element, such as an AWG, it is possible to interrogate the photonic sensors as demonstrated in \cite{Orr2011,Perry2013}. Despite the high sensitivity of this interrogator, special care should be taken to match the spectra of the AWG outputs to the sensors spectra. This might be an issue for integrated sensors such as ring resonators \cite{Peternella2017} since the resonance wavelength, in most of the cases, cannot be predicted during the design due to variations of the fabrication process.

The interrogation method here proposed may be applied to any sensor whose spectrum is finite and is modulated by an external signal. We demonstrate its performance using FBG sensors, but the method is equally suitable to other types of sensors such as ring resonators. To the best of our knowledge, this is the first interrogator based on integrated Fourier Transform (FT) spectroscopy. 
The technique is promising since it benefits from high flexibility,  high sensitivity, and offers a high tolerance to variations of the fabrication process. In the past, FT spectroscopy was applied to demultiplexing FBG sensors \cite{Davis1995,Rochford1999}, but at that time, the speed of the method was limited by the mechanical speed of the mirror. Integrated photonics enables the design of new FT spectrometer implementations. The most common one consists of an array of MZIs with different optical path lengths (OPDs) \cite{Florjanczyk2007,Okamoto2010,Velasco2013, Podmore2017}. Thus, the spectrum can be retrieved by calculating the coefficients of the Fourier cosine series from the interferogram. However, since the number of MZIs is finite, the retrieved spectrum is an approximation to the actual one and a large the number of MZIs is required in order to achieve a high spectral resolution.
 
The design of our integrated FT spectrometer is similar to the one proposed by \cite{Uda2018,Okamoto2013}, where the complex Fourier coefficients of the system are obtained by using 3$\times$3 multi-mode interferometers (MMIs). In our case, however, instead of retrieving the spectrum, we demonstrate that the complex Fourier coefficients can be written as a sum of the individual contributions of the sensors.
We obtain a coupled system of non-linear equations, whose solution gives the modulation of the sensor's resonance wavelength. Since no approximation has been made, the minimum modulation amplitude we experimentally retrieved is 400 fm, more than two orders of magnitude smaller than the spectral resolution of our own FT spectrometer, and limited only by the signal-to-noise ratio of the input signal. Moreover, we demonstrate that the number of interferometers can be as small as the number of sensors, which strongly reduces the device footprint without compromising the interrogator sensitivity. 
Finally, we propose a novel technique for compensating the slow drift with time of the phases of the MZIs due to temperature fluctuations.\cite{Peternella2017,Dandridge2011}. This enables the application of this interrogation method for very low speed photonic sensors. Since the speed is only limited by the electronics, our interrogation method is equally suitable for high speed sensors.

\section{Design and characterization of the FT Spectrometer}

\begin{figure}
\includegraphics[width=0.9\linewidth]{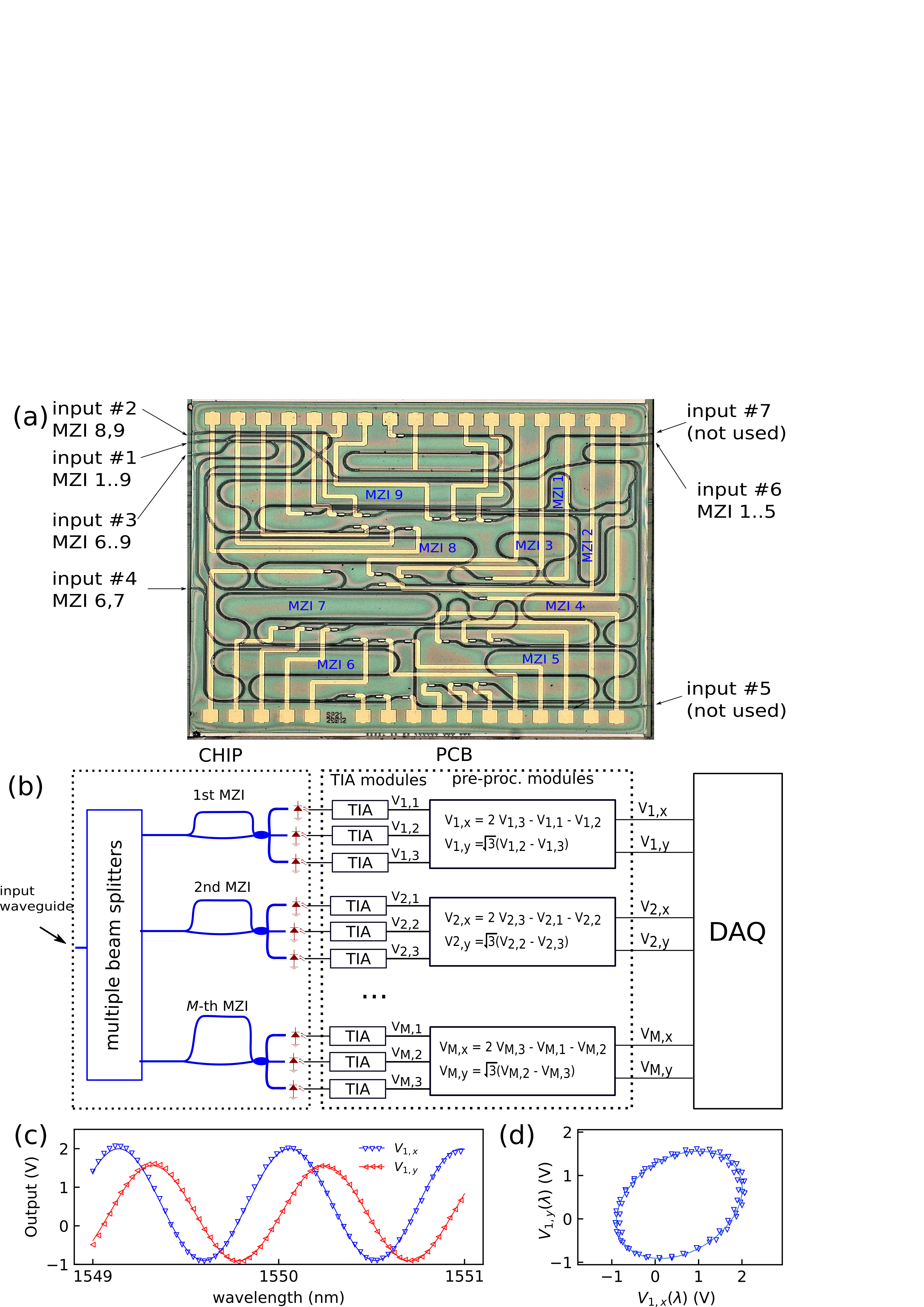}
\caption{  (a) Picture of the FT spectrometer chip. 
$\Delta L_m$ is given by $\Delta L_m = m \Delta L_1$ with $L_1 = 0.710 $~mm, leading to $F_m = F_1/m$, where $m$ is an integer number ranging from 1 to 9. The different MZIs in the figure are identified with the index $m$. (b) Schematic of the FT spectrometer and the PCB that implements the TIAs and a pre-processing module. The outputs are sampled by the DAQ.
(c) Traces of $V_{1,x}$ and $V_{1,y}$ as a function of the laser wavelength. We fitted Eq.~(\ref{eq_VxVy}) against the data points and we obtained $F_1 =$ 921.7~$\pm$~0.5~pm, $\delta \phi_1$ = 17.9 $\pm$0.3$^\circ$, $A_{1,x} = 1.449 \pm 0.003$V and  $A_{1,y} = 1.234 \pm  0.004$V. (d) Lissajous plot of the data points $\left[V_{1,x}(\lambda),V_{1,y}(\lambda)\right]$ of item (d). By fitting an ellipse to the data points we got 1.56~V and 1.09~V for the semi-axis values and 31.2$^\circ$ for the tilt angle with respect to the $x$-axis.
  }
  \label{fig_FT_Spec}
\end{figure}

Fig.~\ref{fig_FT_Spec}a shows a picture of the FT spectrometer. The chip was fabricated by a multi-project wafer run in the Smart Photonics foundry and its dimensions are 4.0 mm $\times$ 4.5 mm.  The chip has a total of 7 inputs, but inputs $\#$5 and $\#$7 are not used, as indicated in the figure. Following the optical path of main entrance (input $\#$1) the light signal is split into nine beams and guided to nine different Mach-Zehnder interferometers.  Other inputs provide access to a limited group of MZIs, allowing the characterization of the sensors using a reduced number of interferometers. For instance, input $\#$6 provides access to MZIs 1-5. 
The chip is glued to a printed circuit board (PCB), to which the chip pads were wire bonded. Outputs per MZI of this PCB were connected to an other PCB which contain three transimpedance amplifiers (TIAs) for the photo-detectors and a pre-processing module. This module gives a linear combination of the outputs, as indicated in the schematic shown Fig.~\ref{fig_FT_Spec}b.

MZIs represent the heart of the on-chip FT spectroscope. All the waveguides have a width of 1.5 $\mu$m and were fabricated in the Deep Etch Layer (see Fig.~6 of \cite{Ciminelli2016}). The 
length difference between the arms range from 0.710 mm to 6.39 mm in steps of 0.710 mm. At the end of the MZI, the light signals from the two arms interfere within a 3$\times$3 MMI (360 $\mu$m length, 11.4 $\mu$m width). 

In this section we characterize the MZIs of the FT spectrometer by considering its response to one particular wavelength $\lambda$. The transmittance for the given wavelength of $l$-th output of the $m$-th MZI is given by:
\begin{equation}
T_{ml}(\lambda) = 
\frac{1}{3}\left[1+v_{ml}\cos\left(2\pi \frac{n_{eff,m}(\lambda) \Delta L_{m}}{\lambda} + \phi_{l}\right)\right],
\label{eq_T_MZI}
\end{equation}
where $v_{ml}$  is the visibility, $n_{eff,m}(\lambda)$ is the effective index of waveguides of the $m$-th MZI, $\Delta L_m$ the arms length difference of the $m$-th MZI, and $\phi_{l}$  is the MZI phase shift given by (120$^\circ$, 0$^\circ$, -120$^\circ$)~for $l$ = 1,2,3 in case the 3$\times$3 coupler is balanced. In our design, the waveguide effective indexes are all the same except by small deviations caused by variations of the fabrication process. Expanding the term $n_{eff,m}(\lambda) / \lambda$ in Taylor series around $\lambda_0$, we obtain:
\begin{equation}
\frac{n_{eff,m}(\lambda)}{\lambda} \cong \frac{n_{eff}(\lambda_0) + n_g + \delta n_{eff,m}}{\lambda_0} - \frac{n_g}{\lambda_0^2}\lambda,
\label{eq_neff_lamb}
\end{equation}
where $\delta n_{eff,m}$ are deviations of the nominal value of the effective index at the $m$-th MZI and $\lambda_0$ a wavelength close to 1550.0~nm. The approximation holds as long as the effect group index ($n_g$) can be considered constant over the spectrum of interest. Replacing Eq.~(\ref{eq_neff_lamb}) in Eq.~(\ref{eq_T_MZI}) we obtain:
\begin{equation}
T_{ml}(\lambda) = \frac{1}{3}\left[
1+v_{ml}\cos\left[2\pi \frac{\lambda}{F_m } - \phi_{l} - \Psi_{m}\right)\right],
\label{eq_T_MZI_expand}
\end{equation}
where $F_m = \lambda_0^2/(n_g \Delta L_m)$ is the free spectral range of the $m$-th interferometer and 
\begin{equation}
\Psi_{m} = \frac{2\pi\Delta L_m}{\lambda_0} \left(n_g+n_{eff}(\lambda_0)+\delta n_{eff,m}\right).
\label{eq_def_psi}
\end{equation}
In our design, $\Delta L_m$ is given by $\Delta L_m = m \Delta L_1$ with $\Delta L_1 = 0.710$~mm, leading to $F_m = F_1/m$, where $m$ is an integer ranging from 1 to 9 and $F_1 = 921.7 \pm 0.5$ pm. $\Psi_{m}$ depends on $n_{eff}(\lambda_0)$, which might change in case of temperature fluctuations, inducing a phase drift in $T_{ml}(\lambda)$. 

Following the schematic of Fig.\ref{fig_FT_Spec}b, it is shown that the outputs of the MZIs are connected to integrated photo-detectors (PD). The PD current $I_{ml}$ is given by $I_{ml}(\lambda) = P_m R_{ph} T_{ml}(\lambda)$, where $P_m$ is the optical power delivered at the $m$-th MZI and $R_{ph}$ is the photodetector responsivity. The outputs of the photo-detectors are send to TIAs, whose outputs voltage are given by:
\begin{equation}
V_{ml}(\lambda) = g_{ml} P_m R_{pm} T_{mk}(\lambda) = \frac{g_{ml} P_m R_{pm}}{3} \left[ 1+v_{ml}\cos\left(2\pi m \frac{\lambda}{F_1 } - \phi_{l} - \Psi_{m}\right) \right],
\end{equation}
where $g_{ml}$ is the transimpedance gain. The 3$\times$3 MMIs were designed to produce interference fringes with similar amplitude and a 120$^\circ$ shift between each other. Aiming for the interrogation of the photonic sensors, the pre-processing module of the PCB combines the TIA output voltages according to \cite{Peternella2017}:
\begin{align}
\label{eq_VxVy}
\begin{split}
V_{m,x}(\lambda) =& 2 V_{m,3} - V_{m,1} - V_{m,2} = A_{m,x} \cos\left(2\pi m \frac{\lambda}{F_1 } - \Psi_{m}\right) + x_{off,m}, \\
V_{m,y}(\lambda) =& \sqrt{3}\left(V_{m,2} - V_{m,3}\right) =  A_{m,y}\sin\left(2\pi m \frac{\lambda}{F_1 } - \Psi_{m} - \delta\phi_m\right) + y_{off,m},
\end{split}
\end{align}
where $V_{m,x}$ and $V_{m,y}$ are 90$^\circ$ phase shift voltages, $A_{m,x}$ and $A_{m,y}$ are the voltage amplitudes, $x_{off,m}$ and $y_{off,m}$ are voltage offsets, and $\delta\phi_m$ is a phase error. If the 3$\times$3 MMI is balanced and the electronic components of the PCB are ideal (ideal operational amplifiers and no variance with respect to the nominal value of the resistors and capacitors), the voltage offsets are zero ($x_{off,m} = y_{off,m} = 0$),
$\delta\phi_m$ = 0, and $A_{m,x} = A_{m,y} = P_m R_{ph} g v$, where the visibility is  $v = v_{m1} = v_{m2} = v_{m3}$ and the TIA gain is $g = g_{m1} = g_{m2} = g_{m3}$.
 In this case, the Lissajous curve $\left[V_{m,x}(\lambda),V_{m,y}(\lambda)\right]$ gives a circle with radius $v P_m R_{ph} g$ centred at the origin.

The transmission spectrum of each MZI has been measured using a tunable laser (Agilent, 81960A). The laser power is set to 6.0 mW and we performed the laser wavelength sweep ranging from 1550~nm to 1551~nm in steps of 1~pm, while the outputs of the pre-processing module are recorded by the digital acquisition module (DAQ, National Instruments, NI 9220). Fig.~\ref{fig_FT_Spec}c shows the measured voltages of the outputs of MZI~1 ($\Delta L=$ 0.710~mm), as well as a fit of the measured data against to Eq.~(\ref{eq_VxVy}). Since $V_{1,x}$ and $V_{1,y}$ have slightly different amplitudes and $\delta\phi_1$ = 17.9$^\circ$, the circle is deformed into a tilted ellipse centred outside of the origin, as shown in Fig.~\ref{fig_FT_Spec}d. Non-idealities and variations of electronic components in the PCB are neglected in the following section, whereas the unbalancement of the 3$\times$3 MMI needs to be considered. In Section 3.2 we discuss how to correct for this.

\section{Interrogation method and experimental setup}

\subsection{The interrogation method}
Here we derive the expressions for determining the resonance wavelengths of the photonic sensors as a function of time. Typically, the spectrum of each sensor has a peaked lineshape, which is modulated by an external signal such as temperature, strain or any other physical or chemical quantity. The photonic sensors are assumed to be wavelength multiplexed. Let there be K sensors with resonance wavelengths $\lambda_k(t)$ at time $t$, where $k=1,...,K$.
The combined spectrum $S(\lambda,\lambda_1(t),...,\lambda_K(t))$ received by the interrogator is given by:
\begin{equation}
S(\lambda,\lambda_1(t),...,\lambda_K(t)) = \sum_{k=1}^K{s_k(\lambda,\lambda_k(t))} = \sum_{k=1}^K{s_k(\lambda - \lambda_k(t))},
\label{eq_def_S}
\end{equation}
where $s_k(\lambda,\lambda_k(t))$ is the spectrum of the $k$-th sensor. 
The signals that are to be sensed induce time dependent modulations of the resonance wavelengths. The resonances $\lambda_k(t)$ must be separated so that the curves $s_k(\lambda,\lambda_k(t))$ do not overlap. In this paper $s_k(\lambda,\lambda_k(t))$ correspond to the reflection spectra of FBGs sensors. However, the method applies also to integrated photonic sensors as the ones described in \cite{Peternella2017}. 

$S(\lambda)$ is assumed to be a poly-chromatic signal and the values of the TIA output voltages are given by:
\begin{equation}
V_{ml}(t) = G \int_{-\infty}^{\infty}{S(\lambda,\lambda_1(t),...,\lambda_K(t))T_{ml}(\lambda) d\lambda},
\end{equation}
where the constant $G$ is given by $G =  (1-\alpha_{c}) g R_{ph}$ with $\alpha_{c}$  the coupling losses. The electronic pre-processing module combines the signals from the three outputs of the interferometers according to Eq.~(\ref{eq_VxVy}), resulting in the two 90$^\circ$ phase shifted voltages $V_{m,x}(t)$ and $V_{m,x}(t)$:
\begin{equation}
V_{m,x}(t) = 3 G \int_{-\infty}^{\infty}{S(\lambda,\lambda_1(t),...,\lambda_K(t))\cos\left(2\pi m \frac{\lambda}{F_1} - \Psi_m\right)d\lambda} + x_{off,m},
\label{eq_Vmx}
\end{equation}
\begin{equation}
V_{m,y}(t) = 3 G \int_{-\infty}^{\infty}{S(\lambda,\lambda_1(t),...,\lambda_K(t))\sin\left(2\pi m \frac{\lambda}{F_1} - \Psi_m\right) d\lambda}  + y_{off,m}.
\label{eq_Vmy}
\end{equation}
As explained in Section 2, the voltage offsets $x_{off,m}$ and $y_{off,m}$ are mainly caused by the fact that the 3$\times$3 MMIs are unbalanced. At the end of a calibration process (see Section 3.2), the offsets are removed by averaging and, at this point, they are neglected.

By defining a complex voltage $\hat{V}_{m}(t) = V_{m,x}(t) + i V_{m,y}(t)$ we obtain:
\begin{equation}
\hat{V}_{m}(t) =  3 G
e^{-i \Psi_m}\int_{-\infty}^{\infty}{S(\lambda,\lambda_1(t),...,\lambda_K(t))\exp\left( i 2\pi \frac{m}{F_1}\lambda \right)d\lambda}.
\label{eq_def_V}
\end{equation}
The chip is characterized after the MZI phase drift has been stabilized, so $\Psi_m$ is constant in time and taken out of the integral in Eq. (\ref{eq_def_V}). In Section 3.3, however, a novel method is presented for compensating the environmental phase drift by using one of the sensors as a reference. 
We assume that $S(\lambda,\lambda_1(t),...,\lambda_K(t))$
vanishes outside the interval $[\lambda_0 - F_1/2,\lambda_0 + F_1/2]$ for 
all times t, where $\lambda_0$ is a wavelength close to 1550.0~nm. 
Then we have:
\begin{equation}
\frac{\hat{V}_{m}(t) e^{i \Psi_m}}{3 G} = 
\int_{\lambda_0 - F_1/2}^{\lambda_0 + F_1/2}{S(\lambda,\lambda_1(t),...,\lambda_K(t))\exp\left( i 2\pi \frac{m}{F_1}\lambda \right)d\lambda}.
\label{eq_V_as_coeff}
\end{equation}
Eq. (\ref{eq_V_as_coeff}) are the Fourier coefficients of the function 
$\lambda\rightarrow S(\lambda,\lambda_1(t),...,\lambda_K(t))$ when considered as periodic function with period $F_1$. This implies that:
\begin{align}\label{eq_FT_series}
\begin{split}
S(\lambda,\lambda_1(t),...,\lambda_K(t)) =& \frac{1}{3 G}\sum_{m=-\infty}^{\infty}{\hat{V}_{m}(t) e^{i \Psi_m}\exp\left( -i 2\pi \frac{m}{F_1}\lambda \right)} \\ =& \frac{2}{3 G}\sum_{m=0}^{\infty}{\left[V_{m,x}(t) \cos\left( 2\pi \frac{m}{F_1}\lambda - \Psi_m \right) - V_{m,y}(t) \sin\left( 2\pi \frac{m}{F_1}\lambda - \Psi_m \right)\right]},
\end{split}
\end{align}
where $\hat{V}_{-m}(t) = \hat{V}_{-m}^*(t)$ since $S(\lambda,\lambda_1(t),...,\lambda_K(t))$ is real.  The chip contains a finite number of $M = 9$ interferometers. The retrieved spectrum $S_M(\lambda,\lambda_1(t),...,\lambda_K(t))$ is given by:
\begin{equation}
S_M(\lambda,\lambda_1(t),...,\lambda_K(t)) = \frac{2}{3 G}\sum_{m=0}^{M}{\left[V_{m,x}(t) \cos\left( 2\pi \frac{m}{F_1}\lambda - \Psi_m \right) - V_{m,y}(t) \sin\left( 2\pi \frac{m}{F_1}\lambda - \Psi_m \right)\right]}.
\label{eq_standard_FT}
\end{equation}
Function $S(\lambda,\lambda_1(t),...,\lambda_K(t))$ differs from $S_M(\lambda,\lambda_1(t),...,\lambda_K(t))$ by the fact that the last one features a finite spectral resolution $\delta\lambda_{res}$ given by:
\begin{equation}
\delta\lambda_{res} = \frac{F_1}{2 M}.
\label{eq_spec_res}
\end{equation}
For $M = 9$, $\delta\lambda_{res} = 50$~pm. Moreover, $S_M(\lambda,\lambda_1(t),...,\lambda_K(t))$ is periodic with period $F_1$. For a large number of interferometers ($M >> K$), $S_M(\lambda,\lambda_1(t),...,\lambda_K(t))$ gives a good approximation to $S(\lambda,\lambda_1(t),...,\lambda_K(t))$ and it is possible to obtain the resonance wavelengths by tracking the peaks of $S_M(\lambda,\lambda_1(t),...,\lambda_K(t))$. However, $\delta\lambda_{res}$ represents a limitation to the minimum modulation amplitude to be experimentally obtained.

In order to determine $\lambda_k(t)$ with higher accuracy and using a reduced number of MZIs we derive a 
non-linear system of equations. We assume in this section that $\lambda(t)$ is known at $t = 0$.  Let
\begin{equation}
\lambda_k(t) = \lambda_k(0) + \delta_k(t), 
\label{eq_def_modulation}
\end{equation}
where $\delta_k(t)$ is the modulation of the resonance wavelength of the $k$-th sensor that we aim to determine. By substituting Eq .~(\ref{eq_def_S}) and Eq. (\ref{eq_def_modulation}) into Eq.~(\ref{eq_def_V}), we obtain:
\begin{equation}
\hat{V}_{m}(t) = 
3 G e^{-i \Psi_m} \sum_{k = 1}^{K}{\int_{-\infty}^{\infty}{s_k(\lambda - \lambda_{k}(0) - \delta_{k}(t))\exp\left(i 2\pi \frac{m}{F_1}\lambda \right)d\lambda}}.
\label{eq_sum_sensors}
\end{equation}
The right-hand side of Eq.~(\ref{eq_sum_sensors}) represents the Fourier transform of $s_k(\lambda - \lambda_{k}(0)-\delta_k(t))$ evaluated at $m/F_1$. Using the shift property of the Fourier transformation, Eq. (\ref{eq_sum_sensors}) is rewritten as:
\begin{equation}
\hat{V}_m(t) = 3 G
\sum_{k = 1}^{K}{ \hat{s}_k(m/F_1) \exp\left[i \left(-\Psi_m + 2\pi \frac{m}{F_1}\lambda_k(0) \right)\right] \exp\left( i 2\pi \frac{m}{F_1}\delta_k(t) \right)},
\label{eq_ft_shift}
\end{equation}
where $\hat{s}_k(m/F_1)$ is the Fourier transform of $s_k(\lambda)$. Let
\begin{equation}
a_{mk} = 3 G \hat{s}_k(m/F_1) \exp\left[i \left(-\Psi_m + 2\pi \frac{m}{F_1}\lambda_k(0) \right)\right].
\label{eq_def_a}
\end{equation}
We rewrite Eq.~(\ref{eq_ft_shift}) as:
\begin{equation}
\hat{V}_{m}(t) =
\sum_{k = 1}^{K}{a_{mk} \exp{\left[\left(i \frac{2\pi}{F_1} \delta_k(t)\right)^m\right]}},
\label{eq_system_non_linear}
\end{equation}
for $m = 1,..., M$.
The coefficients $a_{mk}$ are experimentally determined as explained in Section 3.2.
Eq.~(\ref{eq_system_non_linear}) represents an M$\times$K system of non-linear equations to be solved using Newton's method, where $M$ is the number of interferometers and $K$ is the number of sensors. Hence, the number of interferometers must only be at least as large as the number of sensors (i.e. $M>=K$), which means that the footprint of the device can be relatively small. 
In our chip $M$ = 9.  The system is explicitly written in the in Eq.~(\ref{eq_system_matrix}):
\begin{align}\label{eq_system_matrix} 
\begin{split}
\hat{V}_{1}(t) =& a_{11} \exp\left[i\frac{2 \pi \delta_1(t)}{F_1}\right]+ a_{12} \exp\left[i\frac{2 \pi \delta_2(t)}{F_1}\right] +...+a_{1K} \exp\left[i\frac{2 \pi \delta_K(t)}{F_1}\right], \\
\hat{V}_{2}(t) =& a_{21} \exp\left[2i\frac{2 \pi \delta_1(t)}{F_1}\right] + a_{22} \exp\left[2i\frac{2 \pi \delta_2(t)}{F_1}\right] +...+a_{2K} \exp\left[2i\frac{2 \pi \delta_K(t)}{F_1}\right], \\
...\\
\hat{V}_{M}(t) =& a_{M1} \exp\left[Mi\frac{2 \pi \delta_1(t)}{F_1}\right] + a_{M2} \exp\left[Mi\frac{2 \pi \delta_2(t)}{F_1}\right] +...+a_{MK} \exp\left[Mi\frac{2 \pi \delta_K(t)}{F_1}\right].
\end{split}
\end{align}
It can be show that as long as the phases $2\pi \lambda_k(t)/F_1$ (for $k = 1,...,K$) are different and the initial guess for $\{\delta_1(t),...,\delta_K(t)\}$ is close to the actual solution, the Jacobian $\partial \hat{V}_m/ \partial\delta_k$ is not singular and the Eqs. (\ref{eq_system_matrix}) are independent. From Eq. (\ref{eq_def_modulation}), at $t =0$, $\{\delta_1(0),...,\delta_K(0)\} = \{0,...,0\}$. The solution at time $t$ is taken as an initial guess at $t + 1/f_s$, where  $f_s$ is the sampling frequency. This reduces the computational time and assures that the initial guess and the solution are close to each other. The method is also flexible in the sense that the ratio between the arms length difference of the MZIs ($\Delta L_m/\Delta L_1$) does not need to be an integer number, which would cause the $m$ value in Eq.~(\ref{eq_system_non_linear}) to a fractional number. The equations remain independent as long as the $\Delta L_m$ values are different.

Assuming that the FBG sensors spectra have a Lorenzian lineshape, we replace the Fourier transform of $s_k(\lambda)$ into Eq.~(\ref{eq_def_a}):
\begin{equation}
a_{mk} = \frac{3 G s_{k}^{\text{max}}}{2} \exp\left(\frac{- m OPD_1 }{L_{c,k}}\right) \exp\left[i \left(-\Psi_m + 2\pi \frac{m}{F_1}\lambda_k(0) \right)\right],
\label{eq_solve_integral}
\end{equation}
where $s_{k}^{\text{max}}$ is the maximum value of the Lorenzian of the $k$-th, $OPD_1 = n_g \Delta L_1$ is the optical path difference of MZI 1, and $L_{c,k}$ is the cohenrece length given by:
\begin{equation}
L_{c,k} = \frac{\lambda_0^2}{\pi w_k},
\label{eq_Lc}
\end{equation}
where $w_k$ is the full width half maxima (FWHM) of the Lorenzian. The coherence length limits the maximum OPD value which allows interferometric fringes to be experimentally resolved. 
Eq.~(\ref{eq_solve_integral}) shows that $a_{mk}$ becomes very small when the MZI free spectral range is comparable or smaller than the FWHM of $k$-th sensor.  As discussed in Section 4, the MZIs with larger OPDs are not used due to the strong attenuation and the reduced signal-to-noise ratio (SNR).

\subsection{Calibration and experimental determination of the coefficients}

The coefficients $a_{mk}$ are experimentally determined via the following calibration procedure. 
Let $t^{start}_k$ be the instant of time when the calibration of $k$-th sensor starts and $t^{end}_k$ be the instant of time when the calibration ends for the same sensor. During the time interval $t^{start}_k < t < t^{end}_k$, all sensors are kept at rest, while sensor $k$ is excited. In case sensor $k$ is a temperature sensor, heat is applied (as much as possible) during the calibration. If sensor $k$ is a strain sensor, a large stress is applied (as much as possible). According to Eq.~(\ref{eq_system_non_linear}), for a balanced 3$\times$3 MMI, the $m$-th complex voltage $\hat{V}_{m}(t)$ during the time interval $t^{start}_k < t < t^{end}_k$ is given by:
\begin{equation}
\hat{V}_{m}(t) = a_{mk} e^{i 2 \pi \frac{m}{F_1} \delta_k(t)} + \sum_{l \neq k}^K { a_{ml}} = |a_{mk}| e^{i \theta_{mk}(t)} + c_{mk},
\label{eq_ang_def}
\end{equation}
where $\delta_l(t) = 0$ if $l \neq k$ since no excitation is applied to the other sensors and where $c_{mk} = \sum_{l \neq k}^K { a_{ml}}$ and $\theta_{mk}(t)$ is the complex argument of the term $a_{mk} e^{i m 2 \pi \delta_k(t)/F_1}$, given by:
\begin{equation}
\theta_{mk}(t) = m 2\pi \frac{\delta_k(t)}{F_1} + \arg(a_{mk}).
\label{eq_theta}
\end{equation}
The Lissajous curve $\left( \Re\{\hat{V}_{m}(t)\}, \Im\{\hat{V}_{m}(t)\} \right)$ for $t^{start}_k < t < t^{end}_k$ is given by a circular arc:
\begin{align}\label{eq_circ}
\begin{split}
\left. \left( V_{m,x}(t), V_{m,y}(t) \right)\right|_{t^{start}_k < t < t^{end}_k} = \left.\left( \Re\{\hat{V}_{m}(t)\}, \Im\{\hat{V}_{m}(t)\} \right)\right|_{t^{start}_k < t < t^{end}_k}  \\
=\left.\left[ |a_{mk}| \cos \left(\theta_{mk}(t)\right) + \Re\{c_{mk}\},|a_{mk}| \sin \left((\theta_{mk}(t)\right) + \Im\{c_{mk}\} \right]\right|_{t^{start}_k < t < t^{end}_k},
\end{split}
\end{align}
 where $(\Re\{c_{mk}\},\Im\{c_{mk}\})$ defines the arc centre, $|a_{mk}|$ the radius, and $\theta_{mk}(t)$ the instantaneous angle with the real axis.
 
\begin{figure}[!ht]
  \includegraphics[width=\linewidth]{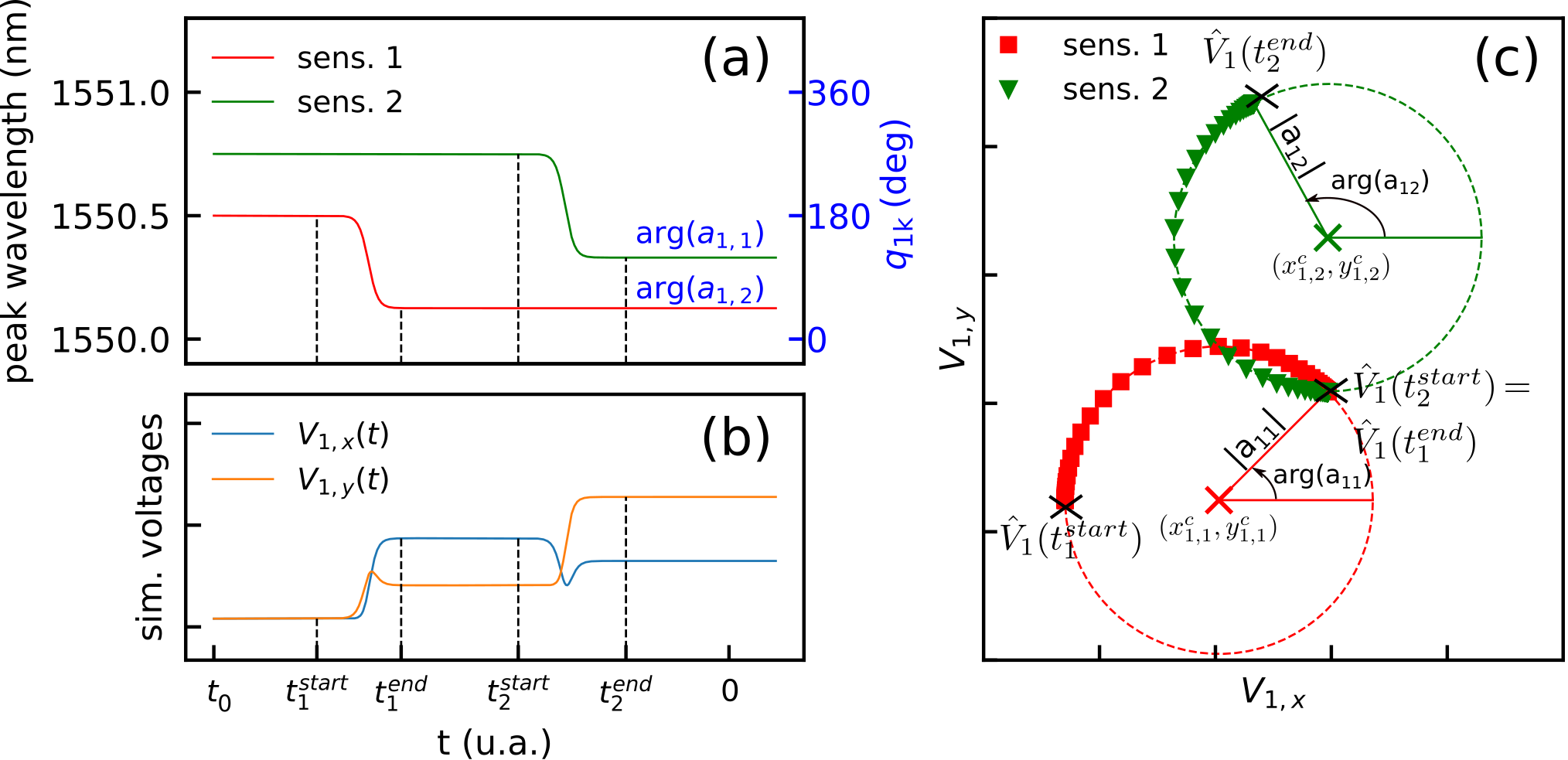}
  \caption{ Illustration of the calibration procedure for two sensors. (a) Independent excitation of sensor 1 and sensor 2. 
  (b) Simulated values of $V_{1,x}(t)$ and $V_{1,y}(t)$ for MZI 1. The changes in time of the functions $V_{1,x}(t)$ and $V_{1,y}(t)$ are caused by the modulation of the peak wavelengths shown in (a).
  The voltages $V_{m,x}(t)'$ and $V_{m,y}(t)'$ (m = 1,...,M) are measured by our acquisition system. $V_{m,x}(t)$ and $V_{m,y}(t)$ are obtained from Eq.~(\ref{eq_correction}). For this simulation, $V_{m,x}(t) = V_{m,x}(t)'$ and $V_{m,y}(t) = V_{m,y}(t)'$.
  (c) Lissajous curve $\left( V_{1,x}(t), V_{1,y}(t) \right)$ for MZI 1. The modulation of the peak wavelength of the sensors induces an angular deflection in the plane of the voltages $V_{1,x}$ and $V_{1,y}$. From the Lissajous curve, the complex modulus and the phase of the coefficients $a_{mk}$ were extracted.
  For this simulation, $F_1=$1.0~nm.
  }
  \label{fig_sim}
\end{figure}

Fig.~\ref{fig_sim} shows a simulation of the calibration for two sensors. The calibration starts at $t = t_0 < 0$ and ends at $t = 0$, when the interrogation procedure starts. During  $t^{start}_1 < t < t^{end}_1$, sensor 2 is kept at rest, while sensor 1 is excited by moving its resonance wavelength from 1550.50~nm to 1550.16~nm, as shown in Fig.~\ref{fig_sim}a. This induces the oscillations of $V_{1,x}(t)$ and $V_{1,y}(t)$ during $t^{start}_1 < t < t^{end}_1$ as shown in Fig.~\ref{fig_sim}b, which are traced as a circular arc in red shown in Fig.~\ref{fig_sim}c. The procedure is repeated for sensor 2: during  $t^{start}_2 < t < t^{end}_2$, while sensor 1 is not excited, sensor 2 changes its resonance from 1550.75~nm to 1550.33~nm. This causes the oscillations from $t^{start}_2 < t < t^{end}_2$ in Fig.~\ref{fig_sim}b which are traced as the circular arc in green shown in Fig.~\ref{fig_sim}c.

As explained in \cite{Peternella2017}, a slight
non-ideal behavior of amplitude and phase of 3$\times$3 couplers are not uncommon and result into a deformation of the circle in an ellipse. 
An ellipse is fitted to the data points $\left( V_{m,x}(t)',V_{y,m}(t)' \right)$ during the interval $t^{start}_k < t < t^{end}_k$,  where $V_{m,x}(t)'$ and $V_{m,y}(t)'$ are the $m$-th MZI voltages measured during the calibration. A larger excitation of the $k$-th sensor results in a larger angular deflection, leading to a more accurate retrieval of geometrical parameters of the ellipse.
The fitting gives the ellipse semi-axis $r_{1,mk}$ and $r_{2,mk}$ (where $r_{1,mk} > r_{2,mk}$), the angle $\alpha$ that represents the rotation of the ellipse with respect to the x-axis, and the ellipse centre $(x^{el}_{mk},y^{el}_{mk})$.  In order to map the ellipse to an circle, the following transformation is applied:
\begin{equation}
\begin{pmatrix}
V_{m,x}(t)  \\ 
V_{m,y}(t)
\end{pmatrix} = 
\begin{pmatrix}
r_{1,mk}/r_{2,mk} & 0\\ 
0 & 1
\end{pmatrix} \begin{pmatrix}
\cos \alpha & \sin \alpha \\ 
-\sin \alpha & \cos \alpha
\end{pmatrix} 
\begin{pmatrix}
V_{m,x}(t)'  \\ 
V_{m,y}(t)' 
\end{pmatrix},
\label{eq_correction}
\end{equation}
where $V_{m,x}$ and $V_{m,y}$ are the corrected values of the 90$^\circ$ phase shifted voltages so that the Lissajous curve $\left(V_{m,x}(t),V_{m,y}(t)\right)$ for $t^{start}_k < t < t^{end}_k$  gives a circle arc with radius $r_{1,mk}$. The correction of Eq.~(\ref{eq_correction}) needs to be performed for all interferometers ($m = 1,...,M$).
Although the ellipse semi-axis $r_{1,mk}$ and $r_{2,mk}$, as well as the corrected radius $r_{1,mk}$ may change according to the sensor (since it depends on its total transmitted or reflected power spectrum) and according to the interferometer (due to the different MZI's coherence lengths), the ellipse eccentricity depends only on the 3$\times$3 MMI, as discussed in Section 2. Thus, for a given interferometer $m$ the ratio $r_{1,mk}/r_{2,mk}$ is constant for $k = 1,...,K$. The design of the 3$\times$3 MMI is the same for all interferometers, hence the ratio $r_{1,m}/r_{2,m}$ is constant for $m = 1,...,M$ as long as the variations of the fabrication process are negligible.

After calculating the 90$^\circ$ phase shifted voltages, the modulus of the coefficients $a_{mk}$ can be obtained. Since the radius of the circle arc obtained for the $m$-th interferometer and the $k$-th sensor is $r_{1,mk}$, the modulus of the coefficients $a_{mk}$, according to Eq. (\ref{eq_circ}), is given by:
\begin{equation}
|a_{mk}| = r_{1,mk}.
\label{eq_mod_a}
\end{equation}
Next, the linear transformation of Eq.~(\ref{eq_correction}) is applied to the point $(x^{el}_{mk},y^{el}_{mk})$, which gives the centre $\left(\Re{\{c_{m,k}\}},\Im{\{c_{m,k}\}}\right)$. The angles $\theta_{mk}(t)$ (for $m = 1,...,M$ and $k = 1,...,K$) are given by:
\begin{equation}
\theta_{mk}(t) = \arctan_2(V_{y,m}(t) - \Im{\{c_{m,k}\}},V_{x,m}(t) - \Re{\{c_{m,k}\}}),
\end{equation}
where $\arctan_2(x,y)$ is the four quadrant arc tangent. 
During the final stage of the calibration of sensor $k$, the angle $\theta_{mk}(t)$ remains constant
because then no excitation is anymore applied to it. By substituting $t = t^{end}_k$ in Eq. ~(\ref{eq_theta}), we obtain:
\begin{equation}
\theta_{mk}(t^{end}_k) = m 2\pi \frac{\delta_k(t^{end}_k)}{F_1} + \arg(a_{mk}) = m 2\pi \frac{\delta_k(0)}{F_1} + \arg(a_{mk}),
\end{equation}
where the calibration procedure ends at $t = 0$. 
According Eq. (\ref{eq_def_modulation}), $\delta_k(0) = 0$. Therefore, the argument of $a_{mk}$ is given by:
\begin{equation}
\arg(a_{mk}) = \theta_{mk}(t^{end}_k) = \theta_{mk}(0).
\label{eq_arg_a}
\end{equation}

The values of $\lambda_k(t)$ (for $k = 1,...,K$) are in general unknown at the end of the calibration ($t = 0$), which contradicts the assumption made in Eq. (\ref{eq_def_modulation}). Here, we refine our previous statement by assuming that the values of $\lambda_k(t)$ are known at $t = t_0$, before the calibration procedure starts. In most of cases, however, the sensors can be calibrated in such a way that their resonance wavelengths return to their initial value at the end of the calibration ($\lambda_k(t_0) = \lambda_k(0)$). In situations where this is not possible (due to a sensor hysteresis, for instance), the values of $\lambda_k(0)$ can be obtained by following the procedure: (a) determine the value of $\delta(t^{start}_k)$ from Eq.~(\ref{eq_theta}) evaluated at $t = t^{start}_k$;  (b) substitute the value of $\delta(t^{start}_k)$ in Eq.~(\ref{eq_def_modulation}) (also evaluated at $t = t^{start}_k$).

After finishing the calibration of all sensors in this way, the offsets are determined by averaging:
\begin{align}\label{eq_remove_offsets}
\begin{split}
x_{off,m} = \frac{1}{|t_0|}\int_{t_0}^{0}{\left\{V_{m,y}(t) - \sum_k{|a_{mk}| \cos\left[ \theta_{mk}(t)\right]}\right\}} dt, \\
y_{off,m} = \frac{1}{|t_0|}\int_{t_0}^{0}{\left\{V_{m,y}(t) - \sum_k{|a_{mk}| \sin\left[ \theta_{mk}(t)\right]}\right\}} dt.
\end{split}
\end{align}
Finally, the complex voltages are computed as function of time to be used in Eqs.~(\ref{eq_system_non_linear}) and (\ref{eq_system_matrix}):
\begin{equation}
\hat{V}_m(t) = \left[V_{x,m}(t) - x_{off,m} \right] + i\left[V_{y,m}(t) - y_{off,m} \right].
\end{equation}

\subsection{Compensation of the phase drift}

Since the effective index in Eq.(\ref{eq_def_psi}) is temperature dependent, local variations of temperature induces the phase $\Psi_m$ to drift. In order to account for this effect, we rewrite Eq. (\ref{eq_def_psi}) according to:
\begin{equation}
\Psi_m(t) = m \frac{2\pi\Delta L}{\lambda_0} \left(n_g+n_{eff}(\lambda_0)(T_0) + \frac{\partial n_{eff}}{\partial T}\Delta T(t)+\delta n_{eff,m}\right) =  \Psi_m(0) + m \Delta\Psi(t),
\label{eq_def_psi_t}
\end{equation}
where
\begin{equation}
\Delta\Psi(t) = \frac{2\pi\Delta L}{\lambda_0} \frac{\partial n_{eff}}{\partial T}\Delta T(t).
\label{eq_def_delta_psi}
\end{equation}
The temperature dependence of the group index $n_g$ and to $\delta n_{neff}$ have been neglected.
Eq. (\ref{eq_def_delta_psi}) indicates that the phases $\Psi_m$ in Eq.~(\ref{eq_ft_shift}),  (\ref{eq_system_non_linear}), and (\ref{eq_system_matrix}) are no longer constant. Eq.~(\ref{eq_ft_shift}) can be rewritten as:
\begin{align}\label{eq_sys_non_linear_phase_drift}
\begin{split}
\hat{V}_m(t) &= 3 G
\sum_{k = 1}^{K}{ \hat{s}_k(m/F_1) \exp\left[i \left(-\Psi_m(0) + 2 \pi \frac{\lambda_{k}(0)}{F_1}\right)\right] \exp\left[ i 2\pi \frac{m}{F_1}\left(\delta_k(t) - \Delta\Psi(t)\frac{F_1}{2\pi}\right) \right]} \\
&=\sum_{m = 1}^{M}{a_{mk}' \exp\left[ \left(i 2\pi\frac{m}{F_1}  \delta_k(t)'\right)^m \right]},
\end{split}
\end{align}
where
\begin{equation}
\delta_k(t)' = \delta_k(t) - \Delta\Psi(t) F_1/(2\pi).
\label{eq_def_delta_psi_prime}
\end{equation}
The right side of Eq.~(\ref{eq_sys_non_linear_phase_drift}) is identical to Eq.~(\ref{eq_system_non_linear}) demonstrating that fluctuations of the environmental phase impacts on the solutions of Eq.~(\ref{eq_system_non_linear}) or Eq.~(\ref{eq_sys_non_linear_phase_drift}). This effect can be corrected by using another sensor as a reference, to which no excitation is applied and its temperature is kept constant. 

Let $\delta_{ref}(t)$ be the solution of Eq.~(\ref{eq_sys_non_linear_phase_drift}) for the reference sensor.
The calibration procedure assures that when the interrogation procedure starts ($t = 0$), the values $\delta_k(0)$ are zero for all sensors ($k = 1,...,K$). Since no excitation 
is applied to the reference sensor, the function $\delta_{ref}(t)$ remains at zero for $t > 0$. Hence, according to Eq.(\ref{eq_def_delta_psi_prime}):
\begin{equation}
\delta_{ref}(t)' = -\Delta\Psi(t) F_1/(2\pi).
\label{eq_def_delta_ref}
\end{equation}
Thus, the phase drift can be compensated by subtracting the term $\Delta\Psi(t) F_1/(2\pi)$ in Eq. (\ref{eq_def_delta_psi_prime}), obtained from Eq.(\ref{eq_def_delta_ref}):
\begin{equation}
\delta_{k}(t) = \delta_{k}(t)' - \delta_{ref}(t)'.
\label{eq_def_delta_compensated}
\end{equation}

\subsection{Experimental setup}

\begin{figure}[!ht]
  \includegraphics[width=\linewidth]{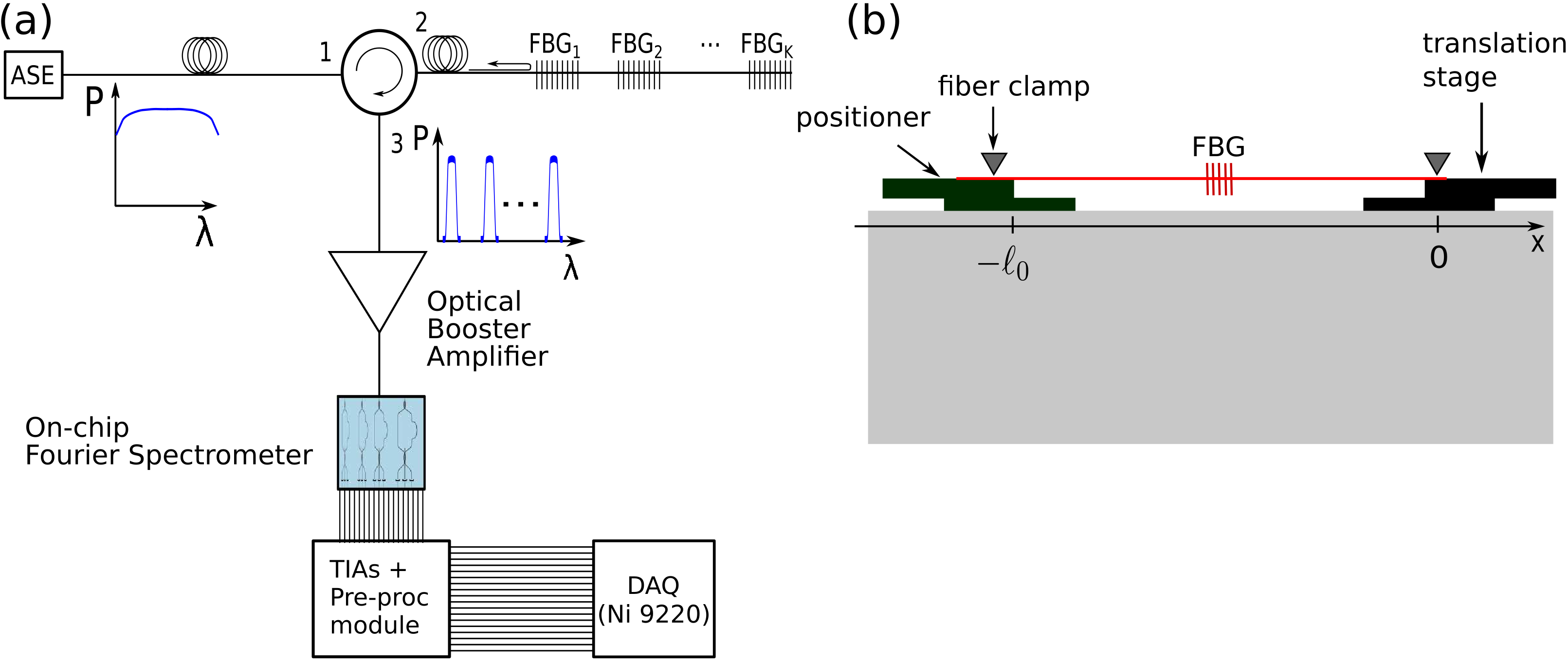}
  \caption{ (a) Schematic of the setup. Light from an ASE source is sent, through a circulator, to the FBG sensor array. The FBG sensors reflect back to the circulator their combined spectrum, which is amplified by an optical booster amplifier (gain = 12dB). Light is coupled to the chip using lensed fibers. In the experiments, we performed the analysis for two cases: using two FBGs and solving Eq.~(\ref{eq_system_matrix}) analytically and using four FBGs and solving Eq.~(\ref{eq_system_non_linear}) numerically. (b) Schematic of the temperature / strain sensors.
  $\ell_0 = 1.74$~m, which is the fiber length between the clamps.}
  \label{fig_setup}
\end{figure}

The schematics of the experiment is depicted in Fig.~\ref{fig_setup}. Light from a broadband amplified spontaneous emission source (ASE, Optolink, OLS15CGB-20-FA) is sent, through a circulator (OZ Optics, FOC-12N-111-9), to the FBG sensor array (Technicasa, T10).  The FBG sensors reflect back to the circulator their combined spectrum, which is amplified by a optical booster amplifier (Thorlabs, S9FC1004P) according to Fig.~\ref{fig_setup}a.  The gain is 12 dB and the light is coupled to the chip using lensed fibers (Oz Optics, TSMJ-3A-1550-9). Outputs of the chip are conveyed to a PCB which implements the transimpedance amplifiers for the photodetectors and an pre-processing module in order to implement Eq.~(\ref{eq_VxVy}) electronically (see  Fig.~\ref{fig_FT_Spec}). The PCB outputs are sampled by the DAQ (National instruments, NI9220), which the maximum sampling speed is 100 kSa/s/channel.

The performance of our interrogator is evaluated using four FBG sensors: three as strain sensors one as a reference sensor, used to compensate the environmental phase drift. The calibration is performed in a such way that $\lambda_{k}(t_0) = \lambda_{k}(0)$. The ends of the fibers containing the FBGs are clamped to the translation stages as shown in Fig.~\ref{fig_setup}b. In order to tune the peak wavelengths $\lambda_{k}(0)$, stress is applied using the manual positioners, avoiding the angles $2\pi \left( \lambda_k(t) \right)/F_1$ to overlap during the experiment. 
FBG~$\#$1 represents the main strain sensor and the translation stage (referred as translation stage 1) to which FBG $\#$1 is attached is controlled by a stepper motor. FBGs~$\#$2 and $\#$3 are the secondary strain sensors and they are both attached to translation stage 2 controlled by another stepper motor. FBG~$\#$4 is the reference sensor and it is attached only to manual positioners.
We programmed the stepper motors to operate in cycles of three steps: (a) the translation stage travels at a constant speed from the position $x = 0$ to $x = \Delta \ell$; (b) The stage rests at $x = \Delta \ell$; (c) The stage returns to the original position. 

Since FBGs~$\#$2 and $\#$3 are secondary strain sensors, we programmed the translation stage to move periodically from the distances $x = 0$ to $x  = \Delta \ell^{(2)} = 30 \mu$m. 
In contrast, the translation stage to which FBG~$\#$1 is attached, travels to different values of $\Delta \ell^{(1)}$ ranging from 0.5 $\mu$m to 200 $\mu$m (these values are shown later in Fig. \ref{fig_strain}). Since the stress to be applied to FBG $\#$1 is much larger compared to FBGs~$\#$2 and $\#$3, the translation stage 1 is programmed to move towards $-x$. Thus, a negative stress applied to FBG $\#$1, avoiding to damage it.
Translation stage 1 repeats three times its motion from $x = 0$ to $x  = \Delta \ell^{(1)}$ and from $x  = \Delta \ell^{(1)}$ to $x = 0$. Thus, the travelling distances $\Delta \ell^{(1)}_{3j+1}$, $\Delta \ell^{(1)}_{3j+2}$ and $\Delta \ell^{(1)}_{3j+3}$ are the same for $j = 0,...,J-1$, where $J$ is the number of different values of $\Delta \ell^{(1)}$.

\section{Experimental results}

As explained in Section 3.4, the performance of our interrogator is evaluated using four FBG sensors: three as strain sensors one as a reference sensor, used to compensate the environmental phase drift. Using manual positioners, a constant stress is applied to all FBGs in such a way that the resonance wavelengths of the sensors are set to $\lambda_{1}(0)$ = 1550.9~nm, $\lambda_{2}(0)$ = 1550.3~nm, $\lambda_{3}(0)$ = 1551.4~nm and  $\lambda_{4}(0)$ = 1549.7~nm.  
The differences of $\lambda_k(t) - \lambda_l(t)$ for $l \neq k$ can be larger than $F_1$ ($F_1$ is the free spectral range of MZI 1) provided that the angles $2 \pi \lambda_k(t)/F_1 \neq 2 \pi \lambda_l(t)/F_1$ for all $l,k = 1...K$. The light signal is coupled to the chip using input $\#$6 (see Fig.~\ref{fig_FT_Spec}a), where the input power is shared among MZIs 1 to 5. Better interrogation results are obtained by sharing the optical power among a reduced number of interferometers since the outputs of the MZIs with larger OPDs are strongly attenuated, according to the discussion in the end of Section 3.1. 

\begin{figure}[!ht]
  \includegraphics[width=\linewidth]{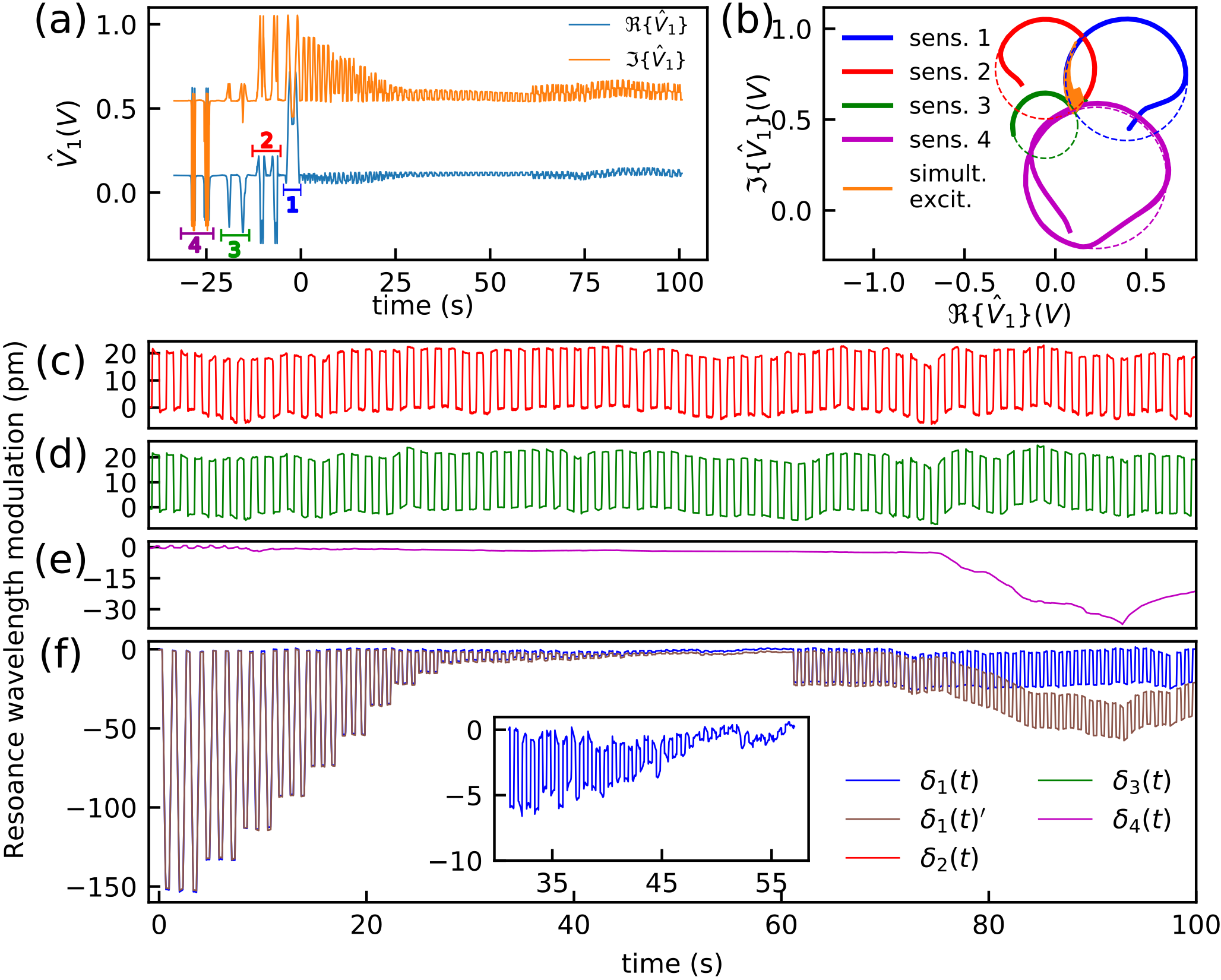}
  \caption{ Main results of the interrogation. (a) Time traces of the real and imaginary parts of $\hat{V}_1 (t)$. A low pass filter (cut-off at 45 Hz) has been applied to the measured voltages $V_{m,x}(t)$ and $V_{m,y}(t)$. The numbers 1,2,3 and 4 indicate the calibration interval ($t_k^{start} < t < t_k^{end}$) for sensors $k = $1,...,4.
  (b) Lissajous plot obtained by plotting the real and imaginary parts of $\hat{V}_1(t)$. During the calibration, the Lissajous curve is a circular arc. During the interrogation, all sensors are simultaneously excited, and an arbitrary Lissajous curve is obtained as shown in orange.
  (c)-(e) Solutions $\delta_2(t)$, $\delta_3(t)$  and $\delta_4(t)'$ of Eq.(\ref{eq_sys_non_linear_phase_drift}) for $t > 0$. FBG $\#$4 is the reference sensor.
  The phase drift was compensated using Eq. (\ref{eq_def_delta_compensated}). (f) Comparison between the solutions $\delta_1(t)$ and $\delta_1(t)'$. The inset shows a zoom of the solution $\delta_1(t)$.}
\label{fig_sol_num}   
\end{figure}

In order to retrieve the coefficients $a_{mk}$, we individually excited the FBG sensors. Following the procedure described in Section 3.2, the complex voltages $\hat{V}_m(t)$ have been obtained by mapping the ellipse arcs to circle arcs according to Eq. (\ref{eq_correction}), and by removing the voltage offsets according to Eq. (\ref{eq_remove_offsets}). Fig.~\ref{fig_sol_num}a shows the real and imaginary parts of $\hat{V}_{1}(t)$, to which a low pass filter (cut-off at 45 Hz) has been applied in order to suppress noise. The real and the imaginary parts of $\hat{V}_1(t)$, shown in Fig.~\ref{fig_sol_num}a, are plotted in Fig.~\ref{fig_sol_num}b as a Lissajous curve. The figure shows four circular arcs, which correspond to the individual excitation of the sensors, obtained from the outputs of MZI $m= 1$ during the calibration. The radii and the angles of the arcs at the end of the calibration procedure give the modulus and argument of the coefficients $a_{mk}$, as described in Section 3.2.

Fig.~\ref{fig_sol_num}b shows, however, that some regions of the Lissajous curve deviate from the expected circular path. This occurs when the resonance wavelengths of two FBGs are about to cross and the spectra of two FBG sensors overlap. This causes that a part of the input optical signal is reflected multiple times in between the FBGs, creating an Fabry-Perot cavity. 
The interference of the electric field which is reflected multiple times between the FBGs leads to the deviations of the circular arcs. To overcome this issue, we followed the calibration described in Section 3.2 using only the parts of the Lissajous curves that are close to circular. For $t > 0$~s, the interrogation starts and the three strain sensors are simultaneously excited. As a result, an arbitrary Lissajous curve is obtained.

Fig.~\ref{fig_sol_num}c-f shows the solution of Eq.~(\ref{eq_system_non_linear}) obtained using the Newton's method. As explained in Section 3.1, the solution obtained at the instant $t$ is used as an initial guess for the Newton's method at the instant $t + 1/f_s$, where $f_s$ is the sampling frequency. As a result, the method converges at any $t$ with a maximum of four interactions. For a sampling rate of 10~kSa/s, about one million of systems of equations needs be solved from $t = $0~s to $t = $~100~s. 
Using an Intel i5-3470 processor, the solution is roughly calculated at a rate of a hundred equations per second and the total computational time is about 2h and 45 min. 

FBGs $\#$2 and $\#$3 are attached to translation stage 2 which periodically travels from $x = 0$ to $x  = \Delta \ell^{(2)} = 30 \mu$m. As a result, the functions $\delta_2(t)$ and $\delta_3(t)$ are time periodic, as shown in Figs.~\ref{fig_sol_num}c and ~\ref{fig_sol_num}d. On the other hand, Fig.~\ref{fig_sol_num}f shows the solution $\delta_1(t)$, which consists of a succession of dips. The dips are obtained because the stepper motor applies a negative stress to FBG $\#$1, as explained in Section 3.4. Since the translation stage repeats its motion three times to a given distance $\Delta \ell^{(1)}$, Fig.~\ref{fig_sol_num}f shows a series of dips grouped by 3 successive ones with approximately the same depth.

\begin{figure}[!ht]
  \includegraphics[width=\linewidth]{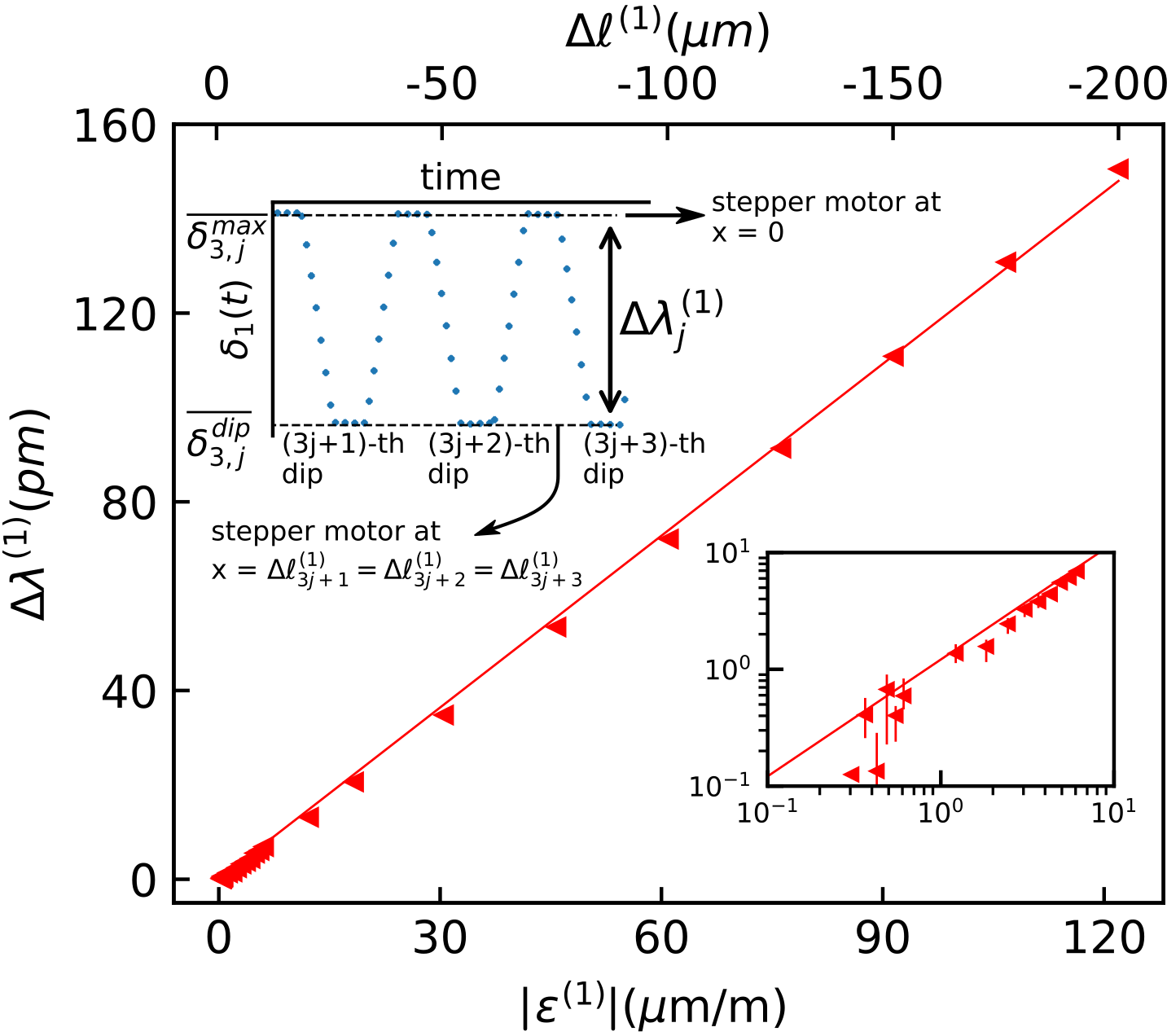}
  \caption{ Modulation amplitude $\Delta\lambda^{(1)}$ of sensor 1 as a function of the strain applied. $\Delta\lambda^{(1)}$ is calculated from  as $\Delta\lambda^{(1)} = \left|\overline{\delta^{dip}_{1,3j}} - \overline{\delta^{max}_{1,3j}} \right|$, where $\overline{\delta^{dip}_{1,3j}}$ and $\overline{\delta^{max}_{1,3j}}$ are defined in upper inset of the figure. A straight line has been fitted to the data points $(|\varepsilon^{(1)}_j|,\Delta\lambda^{(1)}_j)$. The slope, whose value is 1.217$\pm$0.006~pm/microstrain, gives the sensitivity of FBG $\#$1. The inset in the bottom of the figure shows the data points $(\varepsilon^{(1)}_j,\Delta\lambda^{(1)}_j)$ and the straight line fitted in a Loglog plot. The minimum amplitude modulation retrieved is 400$\pm$200 pm. }
\label{fig_strain}   
\end{figure}

Fig.~\ref{fig_strain} shows the modulation amplitude $\Delta\lambda^{(1)}$ for sensor 1 as a function of the strain applied to FBG $\#$1. The strain is assumed to be constant along the fiber and it is defined as:
\begin{equation}
\varepsilon^{(1)}_j = \frac{\Delta \ell^{(1)}_{3j}}{\ell_0},
\label{eq_def_strain}
\end{equation}
where $\varepsilon^{(1)}_j$ is the strain at FBG $\#$1 and $\ell_0$ the fiber length defined in Fig.~\ref{fig_setup}b. The index $3j$ in Eq.~(\ref{eq_def_strain}) appears since the distances $\Delta \ell^{(1)}_{3j}$, $\Delta \ell^{(1)}_{3j+1}$ and $\Delta \ell^{(1)}_{3j+2}$ are the same, as explained in Section 3.4. On the other hand, the modulation amplitude is defined as: 
\begin{equation}
\Delta\lambda^{(1)}_j = \left|\overline{\delta^{\text{dip}}_{1,3j}} - \overline{\delta^{\text{max}}_{1,3j}}\right|,
\end{equation}
where $\overline{\delta^{\text{dip}}_{1,3j}}$ is the time average of function $\delta_1(t)$ at the three adjacent dips $(3j+1)$, $(3j+2)$ and $(3j+3)$, as indicated in the upper inset of Fig.~\ref{fig_strain}. Similarly, $\overline{\delta^{\text{max}}_{1,3j}}$ is the time average of function $\delta_1(t)$ at its $(3j+1)$-th, $(3j+2)$-th and $(3j+3)$-th maxima, which occur when the translation stage rests around the original position $x = 0$. The ratio between the amplitude modulation and the strain gives the sensitivity $S^{(1)}$ of FBG $\#$1:
\begin{equation}
S^{(1)} = \frac{\partial \Delta\lambda^{(1)}}{\partial \varepsilon^{(1)}}.
\end{equation}
By fitting a straight line to the data points $(\Delta\lambda^{(1)}_j,\varepsilon^{(1)}_j)$, we retrieved  $S^{(1)} = 1.217\pm0.006$~pm/$\mu$strain, which agrees with the nominal sensitivity of 1.2~pm/$\mu$strain provided by the manufacturer (Technicasa, T10). The minimum retrieved strain is 365 nanostrain and the corresponding minimum modulation amplitude obtained is $\Delta\lambda_{min}$ = 400$\pm$200~fm. This value is more than two orders of magnitude smaller than the resolution of the FT-spectrometer (50~pm). The value of $\Delta\lambda_{min}$, experimentally retrieved, is not limited by the resolution the FT-spectrometer but only by the SNR of the input signal.

FBG $\#$4 has been taken as a reference sensor and no external excitation is applied to it after the end of the calibration. However, for $t < 10$~s, Fig. \ref{fig_sol_num}d shows small fluctuations of function $\delta_4(t)$ (of the order of a few pm), caused by the cross-talk among sensors. Since the modulation amplitude of FBG $\#$1 is the larger for $t < 10$~s, its cross-talk with FBG $\#$4 is dominant. The maximum cross talk between FBGs $\#$4 and FBGs $\#$1 is about 1$\%$ of the $\delta_1(t)$ value, which is acceptable in most applications.

For $t > 60$~s, the chip is heated up using a Peltier element. The temperature increases about 0.3~$^\circ$C in the chip, a value which is comparable to temperature fluctuations in a temperature controlled room. As explained in Section 3.3, the solution $\delta_4(t)'$ shown in Fig.~\ref{fig_sol_num}e is proportional to the drift of the phase $\Psi_m(t)$. Fig.~\ref{fig_sol_num}f shows a comparison between the solutions $\delta_1(t)'$ and $\delta_1(t)$, this last one obtained using Eq.~(\ref{eq_def_delta_compensated}) (solutions $\delta_2(t)'$ and $\delta_3(t)'$ are not shown). 92.0$\%$ of the phase drift has been compensated. For the sensors presented here, the phase drift influence could have been removed by applying a high pass filter to $\delta_1(t)'$, $\delta_2(t)'$ and $\delta_3(t)'$. However, for low speed sensors such as biochemical sensors \cite{Vos2009}, filtering is not possible since the speed of the sensor is comparable to the phase drift speed. 

Although the method can be applied to high speed sensors, its real time implementation is challenging. On one hand, the speed of the FT-spectrometer is limited only by the electronics and the integrated photo-detectors may respond at frequencies higher than 5 GHz. On the other hand, a system of non-linear equations need to be solved at each instant of time. The computational costs, however, can be reduced by calculating the inverse of the Jacobian  $\partial \hat{V}_m/ \partial\delta_k$ analytically. Using the transformation $z_k(t) = 2\pi (\lambda_k(0) + \delta_k(t))/F_1$, it can be shown that the Jacobian is given by a product of a diagonal matrix and the Vandermond matrix $V(z_k)$. Since analytic expressions do exist\cite{Soto_Equibar2011} for the inverse of $V(z_k)$, the computational time is mainly governed by the time of calculating product of matrices. Moreover, the reduced number of interactions of Newton's method also contributes in reducing the computational time. Nevertheless, the real time interrogation of high speed sensors may require the usage of an application specific computational solution.

\section{Conclusion}

A novel interrogation method based on FT spectroscopy is presented. The technique is promising due to its high flexibility, high sensitivity and reduced interrogator footprint. It can be applied in different situations, in particular, for arrays of integrated sensors where the resonance wavelengths cannot be predicted during the design stage. Three conditions have been identified for the proper interrogation of the sensors: (a) the number of interferometers must only be at least as large as the number of sensors, allowing the interrogator footprint to be relatively small; (b) the MZIs must have different OPDs; (c) the phases $2\pi \lambda_k(t) /F_1$ (for $k = 1,...,K$) needs to be different at any time. If the maximum amplitude modulation of the sensors is known, condition (c) is usually not an issue for FBG sensors, since the Bragg wavelengths could be chosen with an accuracy better than 1.0~nm. In case of integrated ring resonators, it is possible in most situations to design rings with a slightly different lengths, assuring a similar free spectral range, but different resonances. Since the phases depend on $F_1$, the proper design of the FT spectrometer gives an extra flexibility to avoid the phases $2\pi \lambda_k(t) /F_1$ to overlap.

It has been shown that the minimum modulation amplitude experimentally retrieved is not limited by resolution of the FT-spectrometer, but limited only by the signal-to-noise ratio of the input signal. The minimum modulation amplitude obtained is 400 $\pm$ 200 fm and the cross-talk, which also depends on the SNR, is about 1$\%$. Moreover, the phase drift of the interrogator, caused by temperature fluctuations, can be compensated by using one of the sensors as reference sensor to which no external excitation is applied. This is important for low speed sensors where the thermal induced drift of MZI phases is comparable to the speed of the sensors. Our method can also be applied for high speed sensors, but the implementation of real time interrogators require the analytic calculation of the inverse of the Jacobian matrix used in Newton's method. For real time interrogation the the usage of application specific computational solutions may be needed.

\section*{Funding}
F.G.P. is funded by Brazilian Council for Scientific and Technological Development (CNPq).

\section*{Acknowledgement}
We thank Smart Photonics for fabricating the InP chips and COBRA for providing the design of advanced waveguides. We also thank Optocap and Alter Technology Group for performing the wire bonding of our chips to the PCB.


\begin{thebibliography}{25}

\bibitem{Peternella2017}
F.~G. Peternella, B.~Ouyang, R. Horsten, M.~Haverdings, P.~Kat,
  and J.~Caro.
\newblock Interrogation of a ring-resonator ultrasound sensor using a fiber
  mach-zehnder interferometer.
\newblock {\em Optics Express}, 25(25), p. 31622--31639, 2017.

\bibitem{Zhang2015}
C.~Zhang, S.~Liang Chen, T.~ Ling, and L.~J Guo.
\newblock Imprinted polymer microrings as high performance ultrasound detectors
  in photoacoustic imaging.
\newblock {\em Journal of Lightwave Technology}, 33(99), p. 4318--4328, 2015.

\bibitem{Hallynck2012}
E.~Hallynck and P.~Bienstman.
\newblock {Integrated optical pressure sensors in silicon-on-insulator}.
\newblock {\em IEEE Photonics Journal}, 4(2), p. 443--450, 2012.

\bibitem{Vos2009}
K. de~Vos, J. Girones, S. Popelka, E. Schacht, R. Baets, and
  P. Bienstman.
\newblock {SOI} optical microring resonator with poly (ethylene glycol) polymer
  brush for label-free biosensor applications.
\newblock {\em Biosensors and Bioelectronics}, 24(8), p. 2528--2533, 2009.

\bibitem{Hou2017}
X. Zhou, Y. Dai, J.~M. Karanja, F. Liu, and M. Yang.
\newblock Microstructured {FBG} hydrogen sensor based on pt-loaded wo 3.
\newblock {\em Optics Express}, 25(8), p. 8777--8786, 2017.

\bibitem{Liang2018}
Q. Liang, K. Zou, J. Long, J. Jin, D. Zhang, G.
  Coppola, W. Sun, Y. Wang, and Y. Ge.
\newblock Multi-component {FBG}-based force sensing systems by comparison with
  other sensing technologies : A review.
\newblock {\em IEEE Sensors Journal}, 18(18), p. 7345--7357, 2018.

\bibitem{Kersey1996}
A.~D. Kersey.
\newblock {A Review of Recent Developments in Fiber Optic Sensor Technology}.
\newblock {\em Optical Fiber Technology}, 2, p. 291--317, 1996.

\bibitem{Ongqiang2017}
H. Li, X. Ma, B. Cui, Y. Wang, C. Zhang, J. Zhao,
  Z. Zhang, C. Tang, and E. Li.
\newblock Chip-scale demonstration of hybrid {III - V} / silicon photonic
  integration for an {FBG} interrogator.
\newblock {\em Optica}, 4(7), p. 692--700, 2017.

\bibitem{Pustakhod2016}
D. Pustakhod, E. Kleijn, K. Williams, and X. Leijtens.
\newblock High-resolution awg-based fiber.
\newblock {\em IEEE Sensors Journal}, 28(20), p. 2203--2206, 2016.

\bibitem{Yebo2011}
N.~Adello Y., W. Bogaerts, Z. Hens, and R. Baets.
\newblock On-chip arrayed waveguide grating interrogated silicon-on-insulator
  microring resonator-based gas sensor.
\newblock {\em IEEE Photonics Technology Letters}, 23(20), p. 1505--1507, 2011.

\bibitem{Guo2013}
H. Guo, G. Xiao, N. Mrad, and J. Yao.
\newblock Echelle diffractive grating based wavelength interrogator for
  potential aerospace applications.
\newblock {\em Journal of Lightwave Technology}, 31(13), p. 2099--2105, 2013.

\bibitem{Tiwari2013}
U. Tiwari, K. Thyagarajan, M.~R. S., and S.~C. Jain.
\newblock {EDF}-based edge-filter interrogation scheme for fbg sensors.
\newblock {\em IEEE Sensors Journal}, 13(4), p. 1315--1319, 2013.

\bibitem{Passaro2012}
V. M~N Passaro, A.~V. Tsarev, and F. {De Leonardis}.
\newblock Wavelength interrogator for optical sensors based on a novel
  thermo-optic tunable filter in {SOI}.
\newblock {\em Journal of Lightwave Technology}, 30(13), p. 2143--2150, 2012.

\bibitem{Orr2011}
P. Orr and P. Niewczas.
\newblock {High-speed, solid state, interferometric interrogator and
  multiplexer for fiber Bragg grating sensors}.
\newblock {\em Journal of Lightwave Technology}, 29(22), p. 3387--3392, 2011.

\bibitem{Perry2013}
M. Perry, P. Orr, P. Niewczas, and M. Johnston.
\newblock High-speed interferometric fbg interrogator with dynamic and absolute
  wavelength measurement capability.
\newblock {\em Journal of Lightwave Technology}, 31(17), p. 2897--2903, 2013.

\bibitem{Davis1995}
M.~A. Davis and A.~D. Kersey.
\newblock {Application of a Fiber Fourier Transform Spectrometer to the
  Detection of Wavelength-Encoded Signals from Bragg Grating Sensors}.
\newblock {\em Journal of Lightwave Technology}, 13(7), p. 1289--1295, 1995.

\bibitem{Rochford1999}
K.~B. Rochford and S.~D. Dyer.
\newblock {Demultiplexing of interferometrically interrogated fiber Bragg
  grating sensors using Hilbert transform processing}.
\newblock {\em Journal of Lightwave Technology}, 17(5), p. 831--836, 1999.

\bibitem{Florjanczyk2007}
M.w Florja{\'{n}}czyk, P. Cheben, S. Janz, A. Scott, B.
  Solheim, and D.~X. Xu.
\newblock {Multiaperture planar waveguide spectrometer formed by arrayed
  Mach-Zehnder interferometers}.
\newblock {\em Optics express}, 15(26), p. 18176--18189, 2007.

\bibitem{Okamoto2010}
K. Okamoto, H. Aoyagi, and K. Takada.
\newblock {Fabrication of Fourier-transform, integrated-optic spatial
  heterodyne spectrometer on silica-based planar waveguide.}
\newblock {\em Optics letters}, 35(12), p. 2103--2105, 2010.

\bibitem{Velasco2013}
A.~V Velasco, P. Cheben, P.~J Bock, A. Del{\^{a}}ge, J.~H
  Schmid, J. Lapointe, S. Janz, M.~L Calvo, D. Xu,
  M. Florja{\'{n}}czyk, and M. Vachon.
\newblock {High-resolution Fourier-transform spectrometer chip with
  microphotonic silicon spiral waveguides}.
\newblock {\em Optics letters}, 38(5), p. 706--708, 2013.

\bibitem{Podmore2017}
H. Podmore, A. Scott, P. Cheben, A.~V. Velasco, J.~H. Schmid,
  M. Vachon, and R. Lee.
\newblock {Demonstration of a compressive-sensing Fourier-transform on-chip
  spectrometer}.
\newblock {\em Optics letters}, 42(7), p. 1440--1443, 2017.

\bibitem{Uda2018}
R. Uda, K. Yamaguchi, K. Takada, and K. Okamoto.
\newblock Fabrication of a silica-based complex fourier- transform
  integrated-optic spatial heterodyne spectrometer incorporating 120o optical
  hybrid couplers.
\newblock {\em Applied Optics}, 57(14), p. 3781--3787, 2018.

\bibitem{Okamoto2013}
K. Okamoto.
\newblock Fourier-transform, integrated-optic spatial heterodyne (fish)
  spectrometers on planar lightwave circuits.
\newblock In {\em International Conference on Fibre Optics and Photonics}, page
  M2A.1. Optical Society of America, 2012.

\bibitem{Dandridge2011}
A.~Dandridge.
\newblock Fiber optic sensors based on the mach-zehnder and michelson
  interferometers.
\newblock In Eric Udd and William B.~Spillman Jr., editors, {\em Fiber Optic
  Sensors: An Introduction for engineers and scientists}, pages 231--275. John
  Wiley and Sons, Inc., 1991.

\bibitem{Ciminelli2016}
C. Ciminelli, D. D'Agostino, G. Carnicella, F.
  Dell'Olio, D. Conteduca, H. P M~M Ambrosius, M.~K. Smit, and
  M.~N. Armenise.
\newblock {A high-Q InP resonant angular velocity sensor for a monolithically
  integrated optical gyroscope}.
\newblock {\em IEEE Photonics Journal}, 8(1), 2016.

\bibitem{Soto_Equibar2011} F. Soto-Eguibar, and H. Moya-Cessa.
\newblock {Inverse of the Vandermonde and Vandermonde confluent matrices}.
\newblock {\em Applied Mathematics and Information Sciences}, 5(3), p. 361--366, 2011.

\end{thebibliography}
\end{document}